\documentclass[pmlr]{jmlr}

\usepackage{longtable}

\usepackage{booktabs}

\usepackage[load-configurations=version-1]{siunitx} 

\theorembodyfont{\upshape}
\theoremheaderfont{\scshape}
\theorempostheader{:}
\theoremsep{\newline}

\jmlrvolume{1}
\jmlryear{2024}
\jmlrworkshop{AAAI 2024 Workshop AI4ED}

\title[Using LLMs to Assess Tutors on Reacting to Students' Errors]{Using Large Language Models to Assess Tutors' Performance in Reacting to Students Making Math Errors
}

 \author{\Name{Sanjit Kakarla} \Email{Sanjit.Kakarla@gmail.com\\
 \Name{Danielle R. Thomas} \Email{drthomas@cmu.edu}\\
 \Name{Jionghao Lin} \Email{jionghao@cmu.edu}\\
 \Name{Shivang Gupta} \Email{Shivang@cmu.edu}\\
   \Name{Kenneth R. Koedinger} \Email{koedinger@cmu.edu}\\
   \addr Human-Computer Interaction Institute \\Carnegie Mellon University\\5000 Forbes Ave.\\Pittsburgh, PA 15213, USA}}
\begin{document}
\maketitle
\begin{abstract}
Research suggests that tutors should adopt a strategic approach when addressing math errors made by low-efficacy students. Rather than drawing direct attention to the error, tutors should guide the students to identify and correct their mistakes on their own. While tutor lessons have introduced this pedagogical skill, human evaluation of tutors applying this strategy is arduous and time-consuming. Large language models (LLMs) show promise in providing real-time assessment to tutors during their actual tutoring sessions, yet little is known regarding their accuracy in this context. In this study, we investigate the capacity of generative AI to evaluate real-life tutors’ performance in responding to students making math errors. By analyzing 50 real-life tutoring dialogues, we find both GPT-3.5-Turbo and GPT-4 demonstrate proficiency in assessing the criteria related to reacting to students making errors. However, both models exhibit limitations in recognizing instances where the student made an error. Notably, GPT-4 tends to overidentify instances of students making errors, often attributing student uncertainty or inferring potential errors where human evaluators did not. Future work will focus on enhancing generalizability by assessing a larger dataset of dialogues and evaluating learning transfer. Specifically, we will analyze the performance of tutors in real-life scenarios when responding to students' math errors before and after lesson completion on this crucial tutoring skill. 

\end{abstract}

\section{Introduction}
\label{sec:intro}

Personalized tutoring remains a consistently effective academic intervention, significantly benefiting student learning \citep{kraft2021blueprint, reynolds2021prompt}. However, the scarcity of human tutors and lack of essential skills among those available pose a challenge \citep{thomas2023tutor}. Recent advancements in pre-trained large language models (LLMs) offer promise in real-time assessment of tutor performance \citep{chen2022program}. This present work investigates generative AI’s capacity to evaluate tutors’ effectiveness in addressing students’ making math errors. Traditionally, intelligent tutoring systems employing model tracing methods have a reputation for swiftly and directly intervening when students make mistakes \citep{merrill1992effective, merrill1995tutoring, mathan2018fostering}. In contrast, expert human tutors have demonstrated success through subtler, indirect guidance, enabling students to find their own errors and repair them \citep{lepper2002wisdom}. Exploring the assessment of this specific tutoring strategy using LLMs becomes crucial, given its importance in fostering independent error correction among students. 

Recent advances in LLMs offer a host of possibilities aimed to enhance learning, including furnishing explanatory feedback to learners \citep{dai2023can}, with prompt engineering of models leaning towards more of an art than a science \citep{wei2022chain}. Among these, Generative Pre-trained Transformer (GPT) models, particularly the latest iteration, GPT-4, exhibits notable improvements over GPT-3.5. GPT-4 excels in tackling more complex tasks, learns faster, and demonstrates reduced potential for biased or offensive responses \citep{openai}. However, it suffers from slower response and generation times, presenting a notable challenge in handling large transcriptions. Despite its advancements, GPT-4 comes at a considerably higher cost of \$0.03/1K tokens compared to GPT-3.5, which costs \$0.0015/1K tokens for input, marking a 20-fold increase in expense \citep{openai}. Considering the speed and cost-effectiveness of GPT-3.5, our focus lies in evaluating its suitability for assessing tutor performance in practical, real-world tutoring settings. To this end, we aim to investigate the following research questions: \textbf{RQ1}: Can large language models accurately assess components of effective human tutors’ responses to students making errors? \textbf{RQ2}: What is the comparative accuracy and performance between GPT-3.5-Turbo and GPT-4 in assessing tutoring dialogues for how tutors respond to students making errors? 
\section{Related work}
Human tutors are particularly effective when trained on building relationships and fostering rapport with students \citep{marshall2021national}. The online tutor lesson \textit{Reacting to Errors} (Appendix A) is an inspiration for this work \citep{thomas2023tutor}. In this brief, scenario-based lesson, tutors practice responding to a student who has made a math mistake. \textit{Reacting to Errors} and the associated criteria for responding to the student are described.  

\subsection{The \textit{Reacting to Errors} Lesson and Associated Criteria}
In \textit{Reacting to Errors}, a tutor is required to respond to a student’s mistake in solving a math problem according to specified elements of appropriate responses as highlighted in research \citep{lepper2002wisdom, loewenberg2009work}.  These elements encompass praising the attempt or effort; subtly drawing attention to the mistake, and guiding the student toward self-correction \citep{lepper2002wisdom}. Any response that explicitly points out the student’s error, instructs the student on what exactly to do, or simply provides the correct answer is not desirable \citep{lepper1993motivational}. To assess a tutor’s response, we establish five criteria in line with the research-recommended approach. The tutor’s response should be: 1) process- or effort-focused, acknowledging the student’s effort; 2) motivating, avoiding negative language and encouraging the student to recognize the mistake on their own; 3) indirect in addressing the error, using leading questions without negative connotations, such as using words like “mistake” or “error” that may discourage the student; 4) immediate, remaining relevant to the problem at hand; and 5) sincere and accurate, such as ensuring truthful praise to the situation (e.g., praising a student for working hard only when they actually put forth effort) and being mathematically correct. Research suggests that students with high and low self-efficacy measures benefit differently from tutorial dialogue \citep{wiggins2017you}. Low-efficacy students may benefit more from an indirect and more subtle approach when reacting to errors than high-efficacy students, who may not be bothered by tutors calling direct attention to their mistakes. The proposed tutoring strategy is recommended for low-efficacy students, or students who historically struggle in math indicated by a history of low performance and proficiency.      

\subsection{Using Large Language Models (LLMs) to Assess Tutor Moves}
Large Language Models (LLMs) are models trained on and exposed to a variety of information - almost all that exists on the Internet - using artificial neural networks as part of deep learning to process information and create outputs in written language familiar to human text. GPT (Generative Pre-trained Transformer) models, 3.5-Turbo and 4, are examples of LLMs; GPT-3.5-Turbo only accepts text inputs while GPT-4 accepts text alongside image inputs \citep{espejel2023gpt}. An area that captivates researchers is whether these LLMs have the capability to assess particular criteria or performance on highly nuanced and humanistic interactions, such as coaching teachers \citep{wang2023chatgpt} and providing feedback to educators and students \citep{dai2023can, kupor2023measuring}. Past work analyzed the potential of LLMs to guide math tutors in remediating student errors, concluding the best-performing model falls short compared to skilled math teachers \citep{wang2023step}. We expand on past work by: analyzing the ability of generative AI to recognize if an error has been made by the student and, if so, the tutor’s response to the student; and comparing different GPT models to determine cost effectiveness at scale. The purpose of our investigation is to understand if GPT models can assess tutor responses to students. When using LLMs, prompt engineering, or the field of carefully and tactically crafting prompts for generative AI, has a huge impact on the nature of the responses generated \citep{wei2022chain}. Prompt engineering is a paradigm within itself, an area researchers do not fully comprehend, as features of effective prompts and prompting techniques in particular situations are unknown \citep{reynolds2021prompt}. Prompting techniques such as few-shot prompting, where examples are provided, and zero-shot prompting, where no examples are given, are often employed and tested to gauge the model’s responses to eventually develop a prompt that stimulates the model to generate the  responses the user desires \citep{espejel2023gpt}. 

\section{Method}

Initially, we focus on prompting GPT-4 for the creation of synthetic dialogues to calibrate human assessment and determine inter-rater reliability. Although we understand there is no exact substitute for original dialogues from actual tutor-student interactions, we use these synthetic dialogues to determine human graders’ reliability in assessing tutor’s feedback to errors. The synthetic dialogues serve as a proxy for ensuring consistency and reducing bias. We employ GPT-4 to generate 50 tutor-student dialogues prompting the LLM to provide a range in tutor performance. There were 156 words (\textit{SD} = 45.9) and 8.6 utterances (\textit{SD} = 2.7), on average, per dialogue. Appendix B displays the prompt used to generate the synthetic tutoring dialogues. 


\subsection{Human Grader’s Identification of Criteria }
Two human annotators with experience in tutoring assessed tutor performance. As a prerequisite to annotating, graders completed the \textit{Reacting to Errors} lesson and referenced the annotation guide (Appendix C), containing explanations of criteria (Section 2.1): \textit{process-focused}, acknowledging student effort; 2) \textit{motivating}, encouraging the student to find their own mistake; 3) \textit{indirect}, not calling direction attention to the student’s error; 4) \textit{immediate}; and 5) \textit{accurate}, mathematically correct and sincere. If the student did not make a math error, graders coded the dialogue as \textit{no error}. Inter-rater reliability between the two annotators is shown in Table 1  \citep{wan2015kappa}. Appendix D illustrates a synthetic tutoring dialogue that scored five points by both graders.
\begin{table}[htbp]
\centering
\caption{Agreement Among Human Graders}
\label{tab:my-table}
\vspace{10pt}
\begin{tabular}{lccc}
\hline
\multicolumn{1}{c}{\textbf{Criteria}} & \textbf{Agreement Score} & \textbf{Cohen’s Kappa} & \textbf{Interpretation} \\ \hline
\textit{process-focused}                              & 68.8\%                  & 0.59                   & moderate      \\
\textit{motivating}                       & 79.2\%                  & 0.57                   & moderate      \\
\textit{indirect}                     & 73.9\%                  & 0.50                   & moderate      \\
\textit{immediate}                                       & 100.0\%                 & 1.00                   & perfect       \\
\textit{accurate}                               & 91.7\%                  & 0.54                   & moderate          \\ 
\textit{no error}                 & 100.0\%                 & 1.00                   & perfect       \\ \hline
\end{tabular}

\hspace{3cm}

\end{table}

\subsection{Corpus Description, Data Pre-Processing, \& Prompting GPT }
The corpus consists of an unknown number of online tutors (a tutor could be represented in more than one dialogue) who were college students at a Pennsylvanian university. The students were middle school students, ranging from 6th-8th grade, from two schools. The student-level demographics represented in the corpus are unknown, however, the school-level demographics consisted of 52\% Latinx from a California public school and 100\% Black and male from a Pennsylvania charter school. Math proficiency is low at both schools, with  one school at zero percent proficiency suggesting the majority of students have low self-efficacy in learning math. Tutoring was performed remotely using Pencil as a remote communication platform with audio recordings transcribed within the platform. Individual dialogue recordings ranged in size from 100 bytes to 37KB. Transcriptions between 2KB and 8KB were used to provide sufficient utterances to assess dialogue while not overloading and slowing down the processing of  the GPT models. In addition, we strive to keep costs low, particularly when scaling to more transcriptions. For all transcriptions, the tutor was the first utterance in the dialogue, which was used as a guide in diarization. We prompt GPT-3.5-Turbo and GPT-4 using the prompt created (shown in Appendix E), with the temperature at 0. Running the prompt on 50 real-life tutoring dialogues, we  report the absolute performance of each model compared to a human grader. Appendix E displays the prompt used to assess real-life dialogues.

\section{Results and Discussion}
\textbf{Large language models demonstrate proficiency in identifying criteria on how to best respond to students making an error, however, both models exhibit limitations in recognizing if an error was made.} In responding to RQ1 on the ability of LLMs to effectively identify criteria for tutors reacting to students making errors, both GPT-3.5-Turbo and GPT-4 performed fairly well. Table 2 displays the absolute performance of both models in assessing the criteria. GPT-3.5-Turbo and GPT-4 both did particularly well on the criteria of \textit{immediate} and \textit{accurate}, with F1 scores equal to or greater than 0.80 for both these criteria. We posit that these models were able to achieve high levels of accuracy in identifying these criteria as \textit{immediate} was quite straightforward to recognize. The tutor typically discussed situations relevant to the problem, and the \textit{accurate} criteria could be graded by simply ensuring well-established mathematical principles were followed. For the \textit{process-focused} and \textit{indirect} criteria, both models struggled a bit more, as shown by lower F1 scores relative to the rest of the criteria. We believe the models interpreted \textit{indirect} feedback to a student making an error as primarily focused on not giving away the correct response, allowing the terms “mistake” and “error” (e.g., a tutor saying,\textit{“You’re close, but there may be a small error”}), which are phrases our human graders found as discouraging in a tutoring session. The annotation guide for human graders (Appendix C) states, “The tutor avoids the use of words such as \textit{mistake} or \textit{error}.” However, the few-shot LLM prompt does not explicitly state tutors should avoid such terms and only provides examples of met criteria, such as “\textit{You have the right idea}” and “\textit{Explain to me what you did here.}”

Both models encountered challenges in discerning when a student had made an error. Specifically, GPT-3.5-Turbo accurately identified dialogues where the student had not made an error 54\% of the time, while GPT-4 exhibited slightly better performance at 63\%. GPT-4 had 23 usable responses for assessing absolute performance while GPT-3.5-Turbo yielded 17 responses. To evaluate the models’ absolute performance in assessing tutors against the five criteria, calculations were carried out exclusively on the dialogues where both the GPT model and human grader concurred that an error had indeed been made by the student.  

\begin{table}[htbp]
\centering
\caption{Performance Comparison of GPT-3.5-turbo and GPT-4}
\label{tab:my-table}
\vspace{10pt}
\begin{tabular}{lllllll}
\hline

\multicolumn{1}{c}{{\color[HTML]{333333} }}                                    & \multicolumn{3}{c}{{\color[HTML]{333333} \texttt{GPT-3.5-turbo}}}                               & \multicolumn{3}{c}{{\color[HTML]{333333} \texttt{GPT-4}}}                                \\ \cline{2-7} 
\multicolumn{1}{c}{ \textbf{Criteria}} & \multicolumn{1}{c}{Precision} & \multicolumn{1}{c}{Recall} & \multicolumn{1}{c}{F1 Score} & \multicolumn{1}{c}{Precision} & \multicolumn{1}{c}{Recall} & \multicolumn{1}{c}{F1 Score} \\ \hline
\texttt{process-focused}                                                       & 0.46                          & 0.67                       & 0.55                         & 0.57                          & 0.73                       & 0.64                         \\
\texttt{motivating}                                                            & 0.64                          & 0.82                       & 0.72                         & 0.71                          & 0.75                       & 0.73                         \\
\texttt{indirect}                                           & 0.58                          & 0.64                       & 0.61                         & 0.59                          & 0.67                       & 0.63                         \\
\texttt{immediate}                                                             & 0.74                          & 0.88                       & 0.80                         & 1.0                           & 0.87                       & 0.93                         \\ 
\texttt{accurate}                                      & 0.74                          & 0.88                       & 0.80                         & 1.0                           & 0.91                       & 0.95                         \\ \hline
\end{tabular}
\hspace{3cm}

\end{table}

\textbf{GPT-4 performed better in assessing all criteria compared to GPT-3.5-Turbo.} In response to RQ2, the GPT-4 model saw slightly greater performance levels for all five criteria. For criteria such as \texttt{immediate} and \texttt{accurate}, there was an F1 score difference of around 0.10, indicating that GPT-4 was able to more accurately assess the more straightforward criteria. For the rest of the criteria, both models have very similar F1 scores. This raises an important question: is the substantial 20-fold increase in the cost of GPT-4 justified by its enhanced performance compared to GPT-3.5-Turbo (OpenAI, 2023)? We believe the performance of \textbf{GPT-3.5-Turbo is sufficient for our purposes} and upon enhancing the prompt for identifying when an error was made by the student and  \texttt{process-focused} and \texttt{indirect} criteria, we can improve performance. However, despite future improvements, we are unsure if reliance on LLMs alone to evaluate and guide tutors is sufficient, with past work concluding similar results when using LLMs to provide step-by-step remediation of math errors \citep{wang2023step}.

\textbf{GPT-4 shows evidence of inferring and recognizing uncertainty.} For given dialogues where the human indicated no student error was made and GPT-4 said there was an error, we asked GPT-4, \textit{“Point out the area of text where the student made an error.”} GPT-4’s notable responses included \textit{“The error made by the student is not explicitly mentioned…However, it can be inferred from the dialogues…where the student expresses some uncertainty” and “The student is confused about how to rewrite a fraction… ‘Divided by what?’ showing uncertainty.” }These responses show that GPT-4 has the capability of reasoning that the student is confused and has consequently likely made an error. The model also associated uncertainty around a problem as being an error, indicating that it is more likely to indicate situations as errors when compared to a human grader who is more prone to simply analyzing the situation without making any inferences while also understanding when exactly a student is feeling doubtful by understanding the context of the tutoring situation. 

\section{Limitations, Future Work, \& Conclusion}
This investigation contains several limitations. First, the dataset comprises a rather small number of dialogues. Increasing the number of dialogues will enhance the generalizability of our findings. Second, the real-life transcriptions we employ include fragmented conversations with many students inputting responses in the chat or providing non-verbal communication through video. We had the initial goal of keeping our costs low and running a single prompt to assess dialogues. Prompting the LLM to check if an error was not made before assessing the criteria may be beneficial. Third, the few-shot prompt had on average three correct examples of tutor responses per criteria. Including incorrect examples of tutor responses by criteria may clear up the ambiguity on what constitutes an error made by the student (e.g., a student displaying uncertainty). Future work consists of firstly analyzing more transcriptions. Second, we will prioritize the synchronization of chat and audio within real-life dialogues and employ more effective data-cleaning techniques. Third, we plan on employing LLMs to assess other specific tutoring skills \citep{chhabra2022evaluation} with past work demonstrating tutor learning \citep{thomas2023tutor}. Ultimately, assessing evidence of learning transfer from the completion of scenario-based lessons is our goal \citep{macaulay1999transfer}. For example, among tutors who completed the lesson \textit{Reacting to Errors}, do we find evidence of tutors modifying their behavior after completing the lesson? Presently, we have more than 20 online tutor lessons aligned with specific tutoring skills such as \textit{Responding to Negative Self-Talk} and \textit{Determining What Students Know.}

In this study, we investigated the capability of GPT-3.5-Turbo and GPT-4 to identify criteria for how tutors should respond to students making math errors. Our findings indicate that both LLMs perform adequately in assessing certain criteria, but they encounter challenges in accurately determining whether the student has made an error. GPT-4 exhibits a slight advantage over GPT-3.5-Turbo, demonstrating the ability to make inferences and judge uncertainty, however, GPT-3.5-Turbo appears to suffice for scalable dialogue assessment while maintaining cost-efficiency. Future research will involve using a larger dataset, exploring learning transfer, and applying this method across different tutoring skills.       

\bibliography{jmlr-sample}
\appendix

\section{Figure 1: Scenario from Reacting to Errors Lesson}\label{apd:first}

\begin{figure}[htbp] 
    \centering 
    \includegraphics[width=0.9\textwidth]{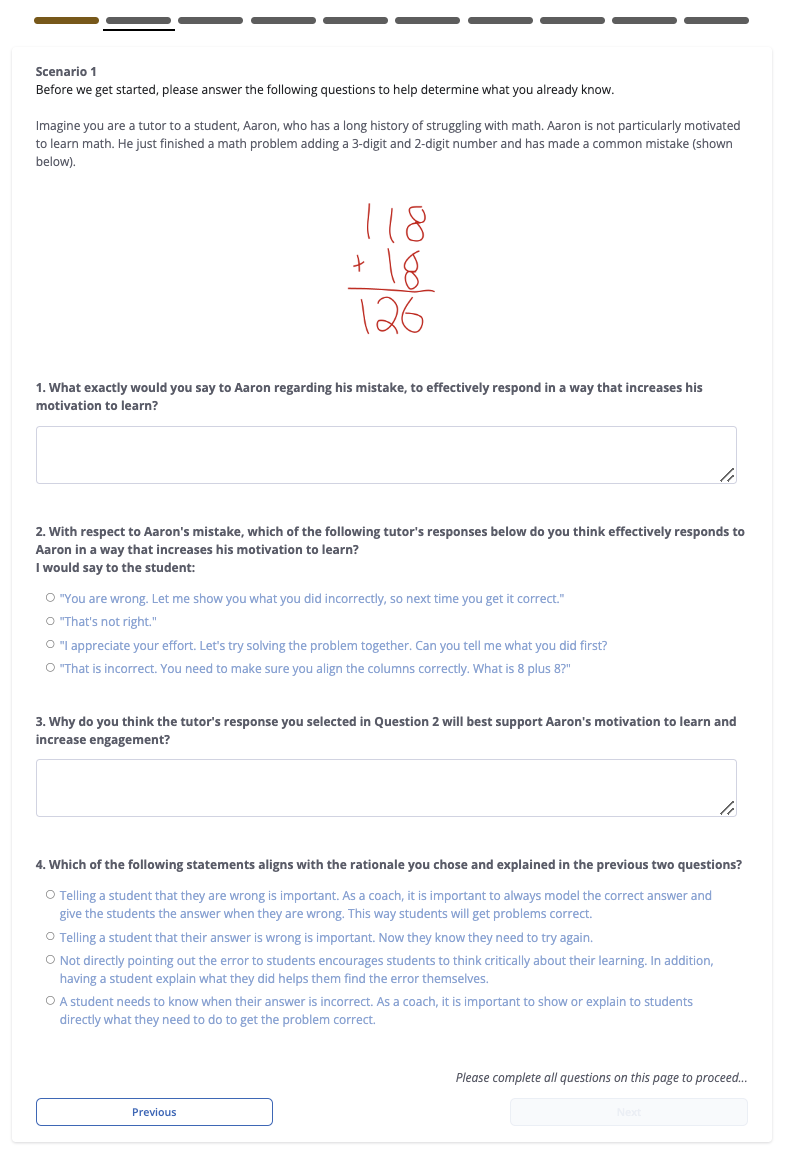} 
    \caption{Scenario from \textit{Reacting to Errors} lesson}
    \label{fig:your_label}
\end{figure}

\section{Prompt to Generate Synthetic Dialogues
}\label{apd:second}

Provide me with a transcript of a tutoring session between a tutor and a middle school student. In this transcript, the student makes a math error (a common math mistake, or forgets to do something). Then, the tutor responds to the student’s error, and they have a conversation. Provide me with 50 tutor and student transcripts with each one approximately 200 words on average. The tutors in each of the transcripts should be of varying experience levels, with performance ranging from poor to very well on how they react to a student making an error.

\section{Annotation Guide}\label{apd:third}

You will see multiple tutor-student interaction transcriptions - both synthetic and authentic. In these transcriptions, a math error may have been made by the student with the tutor, subsequently, reacting to the student’s error. We have provided a description of the five criteria you will use to assess a tutor’s performance in responding to students making an error (Table 3 below). Provided is the coding, along with sample utterances, that meet or do not meet the criteria (Table 4 below). There are five criteria we want you to grade our transcriptions based on: 1). \textit{process-focused}; 2). \textit{motivating}; 3). \textit{indirect}; 4). \textit{immediate}; 5). \textit{accurate}. Part of the grading process will require an understanding of if the student made an error in order to grade the remaining criteria. Let’s dive further into each criterion and what we want you to specifically identify in each transcription. To meet the criteria provided in the table below the tutor may perform any of the following actions for each of the given criteria. 
\begin{table}[]
\caption{Criteria with Explanations/Actions by the Tutor}
\label{tab:my-table}
\vspace{10pt}
\begin{tabular}{ll}
\hline
Criteria                 & Explanation/Action by the Tutor                                   \\ \hline
\textit{process-focused} & \begin{tabular}[c]{@{}l@{}}The tutor:\\ -aims to praise the student’s effort.\\ -acknowledges that the student has made a genuine attempt to solve \\ the problem.\\ -shows that the tutor cares about the student in an educational \\ context.\\ -encourages a growth mindset and builds persistence.\end{tabular}                                                      \\ \hline
\textit{motivating}      & \begin{tabular}[c]{@{}l@{}}The tutor:\\ -avoids negative and demeaning language.\\ -encourages the student to find their mistake on their own.\\ -boosts the confidence of students and builds up their love for math.\end{tabular}                                                                                                                                       \\ \hline
\textit{indirect}        & \begin{tabular}[c]{@{}l@{}}The tutor: \\ -indirectly calls attention to the error made.\\ -avoids the use of words such as “mistake” or “error” with negative, \\ demotivating connotations.\\ -uses leading questions to ensure the student has the ability to find \\ their own mistakes, which allows students to expand their \\ mathematical abilities.\end{tabular} \\ \hline
\textit{immediate}       & \begin{tabular}[c]{@{}l@{}}The tutor:\\ -discusses situations relevant to the problem.\\ -uses specific numbers and refers to particular steps of the problem to\\ explain to the student how an error was made.\\ -allows the student to remain engaged with the tutor.\end{tabular}                                                                                     \\ \hline
\textit{accurate}        & \begin{tabular}[c]{@{}l@{}}The tutor: \\ -gives sincere feedback to the student, while avoiding unnecessary \\ exaggerations of the student’s ability.\\ -gives feedback that is mathematically sound and follows existing \\ mathematical principles/concepts.\\ -provides praise and encouragement that is relevant to the scenario.\end{tabular}                       \\ \hline
\end{tabular}
\end{table}

Now that you have understood the criteria and their importance in a tutoring session for both the tutor and student, let’s look at some examples of tutor utterances that meet these respective criteria along with reasoning. Met criteria receive a score of 1 and unmet criteria receive a score of 0.

\begin{table}[]
\caption{Criteria with Annotated Example Utterances}
\label{tab:my-table}
\vspace{10pt}
\resizebox{\columnwidth}{!}{%
\begin{tabular}{ll}
\hline
Criteria                 & Annotated Example Utterances with Explanation                                                                                                                                                                                                                                                                                                                                                                                                                                                                                                                                                                                                 \\ \hline
\textit{process-focused} & \begin{tabular}[c]{@{}l@{}}Criteria met (1):\\ “I love how you worked hard on that problem!”\\ → Praises the student for their effort and process regardless of whether they got the \\ answer correct.\\ “Really insightful attempt on that step”\\ →  Again, this response focuses on making the student feel confident about their effort.\\ Criteria not met (0):\\ “Good job getting it right!”\\ → Putting focus on getting the question correct rather than praising student effort.\end{tabular}                                                                                                                                      \\ \hline
\textit{motivating}      & \begin{tabular}[c]{@{}l@{}}Criteria met (1): \\ “Keep working hard as you are doing already”\\ →  Encourages the student to continue working hard for the rest of the tutoring session \\ as their previous efforts were met with positive feedback.\\ “You got this!”\\ →  Motivating feedback that will fuel the student to persist past temporary struggles.\\ Criteria not met (0):\\ “You’re not doing a very good job; you’re making so many mistakes.”\\ → Demotivating and discouraging feedback that deteriorates student morale.\end{tabular}                                                                                       \\ \hline
\textit{indirect}        & \begin{tabular}[c]{@{}l@{}}Criteria met (1): \\ “How did you approach this particular step of this problem?”\\ →  Example of a leading question that allows the student to investigate an aspect of a \\ problem where they made a mistake without directly being told by the tutor.\\ “You have the right idea.”\\ →  Using a positive tone to make a student feel like they have the right thought process.\\ Criteria not met (0):\\ “It looks close, but you have made an error.”\\ → Directly stating that an “error” has been made, which can discourage the student from\\  finding their own mistake.\end{tabular}                    \\ \hline
\textit{immediate}       & \begin{tabular}[c]{@{}l@{}}Criteria met (1):\\ “Look back at step 3 of the problem”\\ →  Referring to specific steps in the problem that the student is aware of.\\ “Analyze what you did when subtracting the fractions in step 5”\\ →  Referring to an exact skill and an exact problem.\\ Criteria not met (0):\\ “Yeah, I believe you mentioned your thought process for solving this problem, I don’t \\ know… like last week or something. I’m not sure”\\ → This shows uncertainty and a lack of care while also disengaging the student by briefly\\  attempting to recall information the student likely isn’t sure of.\end{tabular} \\ \hline
\textit{accurate}        & \begin{tabular}[c]{@{}l@{}}Criteria met (1):\\ “Great effort so far” when the student has actually put forth effort\\ →  Shows accuracy in understanding the social context of the tutoring situation.\\ “To add fractions of different denominators, we need a common denominator.”\\ →  Stating a mathematical fact to teach the student how to approach a problem.\\ Criteria not met (0):\\ “Anything divided by zero is equal to 1.”\\ → This is an incorrect and inaccurate statement that goes against foundational \\ mathematical principles.\end{tabular}                                                                           \\ \hline
\end{tabular}%
}
\end{table}

\section{Example of a Synthetic Dialogue}\label{apd:fourth}
Example of a synthetic tutoring dialogue that was graded as a 5 by our human graders:

\textbf{Student}: So, I have the equation 4y + 7 = 31. I subtract 7 from 31 to get 24, then y equals 24.

\textbf{Tutor}: Let's go through this step by step. You subtracted 7 from 31 correctly, but what should you do next?

\textbf{Student}: Isn't y just 24? I subtracted 7 from both sides.

\textbf{Tutor}: You did the subtraction right, but remember, y is being multiplied by 4. So, what should you do with 24?

\textbf{Student}: Oh, do I divide it by 4?

\textbf{Tutor}: Exactly! Can you do that division?

\textbf{Student}: Sure, 24 divided by 4... that's 6. So, y is 6?

\textbf{Tutor}: That’s correct. Why is dividing by 4 the right step here?

\textbf{Student}: Because the original equation was 4 times y, and I need to reverse that to solve for y?

\textbf{Tutor}: Precisely. You're undoing the multiplication to isolate y.

\textbf{Student}: I see my mistake. I forgot about the multiplication part.

\textbf{Tutor}: It's a common error. Always look at what's being done to the variable and do the opposite to isolate it.

\textbf{Student}: Got it. So, I should always reverse the operation on the variable.

\textbf{Tutor}: Yes, that's key in solving equations. Great job understanding it!

\section{Prompt of the LLM}\label{apd:fifth}

\noindent Please score the following tutor-student transcript for five criteria that represent how effective the tutor is in reacting to a middle school student who has made an error in a virtual tutoring session. If the student does not make an error, immediately score the transcript with a -1 and ignore the rest of the prompt. The following are criteria for assessing the tutor:\\
C1) Process or effort-focused-  meaning the tutor praises the student for effort, and not necessarily getting the answer correct. An example of effort-focused praise would be a tutor saying to a student, “I love how you worked hard on that problem,” “Really insightful attempt on that step,” or “Good work on that step”;\\
C2) Motivating- such as encouraging the student by prompting the student to identify their own mistake. An example of motivating the student would be saying, “You’re doing very well,” “ Keep working hard as you are doing already,” or “You got this!”;\\
C3) Indirect attention to the actual error- by using leading questions. An example of being indirect to the actual error would be saying, “You have the right idea”, “Take a closer look” or “Explain to me what you did here”;\\
C4) Immediate, focused on the current scenario- An example of being immediate would be referring to particular steps of the current problem such as saying “Look back at step 3 of the problem” or “Analyze what you did when subtracting the fractions in step 5”;\\
C5) Sincere, truthful, and mathematically accurate- An example of being sincere would be saying, “Great effort so far” only after a student has put forth effort. Being mathematically accurate means the tutor instructs using the correct mathematical principles, for example, saying that, “Area equals length times width” or “To add fractions of different denominators, we need a common denominator.”\\
Score each transcript on each of the 5 criteria independently, providing a score of 0, if the individual criterion is not met, and 1, otherwise. Please provide a separate score of 0 or 1 for each of the five criteria: C1) process or effort-focused; C2) motivating; C3) indirect attention to the actual error; C4) immediate; C5) sincere. If the student does not make a mistake or error in the transcript, please immediately return -1 for all criteria.\\
Transcript Start ---\texttt{<TUTORING TRANSCRIPT>}--- Transcript End.\\
Given the earlier transcript, please return your scores as a JSON file following the format, {C1: 0/1/-1, C2: 0/1/-1, C3: 0/1/-1, C4: 0/1/-1, C5: 0/1/-1}, e.g., {C1: 0, C2: 1, C3: 1, C4: 1, C5: 0}. Your JSON: \texttt{<JSON FILES>}

\end{document}